\documentclass[aps,11pt,preprintnumbers,nofootinbib,floatfix]{revtex4}

\usepackage{graphicx}
\usepackage{epsfig}
\usepackage{dcolumn}
\usepackage{subfigure}
\usepackage{multirow}
\usepackage{amsmath}
\usepackage{amssymb}

\begin{document}

\title{Holographic model with power-law Maxwell field for color superconductivity}


\author{Cao H. Nam}
\email{nam.caohoang@phenikaa-uni.edu.vn}  
\affiliation{Phenikaa Institute for Advanced Study and Faculty of Fundamental Sciences, Phenikaa University, Yen Nghia, Ha Dong, Hanoi 12116, Vietnam}
\date{\today}

\begin{abstract}%
Studying the color superconductivity (CSC) phase is important to understand the physics in the core of the neutron stars which is the only known context where the CSC phase might appear due to the gravitational force squeezing the matter to a sufficiently high density. We propose a simple holographic dual description of the CSC phase transition in the realistic Yang-Mills theory with the power-law Maxwell field. We find the CSC phase transition with the large color number in the deconfinement phase, which is not found in the case of the usual Maxwell field, if the power parameter characterizing for the power-law Maxwell field is sufficiently lower than one but above $1/2$ and the chemical potential is above a critical value. However, the power parameter is not arbitrary below one because when this parameter is sufficiently far away from one it leads to the occurrence of the CSC state in the confinement phase which is not compatible with a nonzero vacuum expectation value of the color nonsinglet operator.
\end{abstract}

\maketitle

\section{Introduction}
In analogy to the electron Cooper pairs in condensed matter physics, at sufficiently high density (or the chemical potential) and low-temperature quarks with the attractive interaction in the color antitriplet channel are expected to form the Cooper pairs (associated with the diquark operator) near the Fermi surface. The quark Cooper pairs are not color singlets and hence their condensation would break spontaneously the $SU(3)_C$ local gauge symmetry of quantum chromodynamics (QCD) by which the gluons acquire a mass (corresponding to color Meissner effects) through the Higgs mechanism where the quark Cooper pairs play the role of the Higgs particles. This phase is well-known as the color superconductivity (CSC) phase or the Higgs phase of QCD, suggested first by the authors in \cite{Collins1975}. For a nice review see \cite{Alford2008}. 

The only known conditions of the density and temperature where the CSC phase might appear are in the core of the neutron stars where the matter density can reach up to ten times of the nuclear saturation density. Therefore, the investigation of the CSC phase is important to understand the physics of neutron stars as well as compact stars. Here, the effects on the observations of the neutron stars caused by the presence of the CSC phase could allow us to resolve the deviations between the experimental measurements and the theoretical predictions.

Due to the asymptotic freedom, it is possible to study the CSC phase transition at very high densities where QCD becomes weakly coupled. Since the condensation pattern can be exactly calculated using the perturbative approach \cite{Son1999}. However, for sufficiently high densities relevant to the real context such as the core of the neutron stars which are the orders of the strong coupling scale, perturbative QCD is not able to be applied although there is the existence of the deconfined quarks. Most QCD investigations in the nonperturbative regime are numerically performed by lattice gauge theory which is difficult due to the sign problem of the fermion determinant.\footnote{Note that, the sign problem can be avoided by restricting the calculations to the imaginary chemical potential \cite{Roberge1986,Bilgici2008,Bilgici2010,Ghoroku2020}.}

In the last two decades, the gauge/gravity duality or the holographic approach has provided a powerful tool to explore the strongly coupled quantum field theories using the weakly coupled gravitational dual theories \cite{Maldacena,Witten,Gubser}. Based on this approach, a bottom-up holographic model has been constructed to study the CSC phase transition in the Yang-Mills (YM) theory \cite{Ghoroku2019} which extended the work in \cite{Fadafan2018} with considering the backreaction of the matter part on the spacetime geometry and the work in \cite{Basu2011} with the complex scalar field. The holographic investigations of the CSC phase transition have lately received attention in the presence of the higher derivative corrections \cite{Nam2021,Fadafan2021} and with including a dilute gas of instantons which is introduced to study the nuclear matter \cite{Ghoroku2021}. 

It was pointed out by the authors in \cite{Ghoroku2019} that the CSC phase transition can occur for $N_c$ (where $N_c$ is the color number of quark) to be small. But, the CSC phase transition could be not found in the case of $N_c\geq2$ and hence this holographic model does not provide a gravitational dual description for the CSC phase transition of the realistic YM theory (i.e. $N_c\geq2$). This restriction can be resolved by considering Einstein-Gauss-Bonnet (EGB) gravity \cite{Nam2021}. Unfortunately, in order to indicate the CSC phase transition for $N_c\geq2$, the magnitude of the GB coupling parameter is required to be rather large, which leads to the violation of the causality bound of the boundary field theory \cite{Camanho2010,Escobedo2010} and is beyond the region of the classical gravity.

The goal of the present work is to construct a holographic model with a power-law Maxwell field which is simple but allows us to investigate the CSC phase transition in the realistic YM theory. It should be noted here that the power-law Maxwell electrodynamics is motivated by the conformal invariance of the action of the $U(1)$ gauge field in arbitrary dimensions \cite{Hassaine2007} and is a type of nonlinear electrodynamics which can lead to the regular black hole solutions \cite{Ayon-Beato98,Ayon-Beato1999a,Ayon-Beato1999b,Bronnikov2001,Dymnikova2004,Schee2015,Nam2018,Singh2018,SGGosh2019,Nam2019} and also are used to study holographic superconductors \cite{Pan2011,Gangopadhyay2012,Lai2015,Salahi2016,Ghorai2016,Jiang2016,Ghazanfari2018,Nam2019b,TZhang2022}. In Sect. \ref{GlSU3C}, we discuss the main obstacle in attempting to study the CSC phase transition based on the holographic approach and indicate a situation where this obstacle can be avoided due to $SU(3)_C$ gauge symmetry of QCD  behaving as just a global symmetry. In Sect. \ref{GDM}, we propose the gravitational dual model of the CSC phase transition which is given by the system of Einstein gravity coupled minimally to the power-law Maxwell field and a charged scalar field in the asymptotically anti-de-Sitter (AdS) spacetime. The action of the system and the equations of motion in the geometric configurations dual to the confinement and deconfinement phases are presented in detail. The CSC phase appears when the quark Cooper pairs have the nonzero vacuum expectation value (VEV) or the scalar field in the gravitational dual theory condenses around the event horizon of the planar AdS black hole. In Sect. \ref{CSCPT}, we point out that if the power parameter (which characterizes for the power-law Maxwell field) is sufficiently smaller than one then the CSC state would appear in the deconfinement phase of the realistic YM theory as the chemical potential is above a critical value. In Sect. \ref{PhaseD}, with the critical line associated with the CSC phase transition found by solving numerically the equations of motion near the critical chemical potential and the free energy of the background configurations from computing the Euclidean on-shell action of the system, we obtain the phase diagram in the $\mu-T$ plane. The summary is given in the final section.

\section{\label{GlSU3C} $SU(3)_C$ color symmetry as a global symmetry}

The CSC phase appears when the diquark operator of the formal form $\langle qq\rangle$ condenses, which leads to the spontaneous breaking of $SU(3)_C$ color symmetry. In fact, the $\langle qq\rangle$ operator is colored or not gauge-invariant and hence this has been the main obstacle for building a holographic model of the CSC phase transition because the bulk fields in the gravity side are always dual to the gauge-invariant operators in the boundary field theory side. However, the authors in Ref. \cite{Fadafan2018} have pointed out a situation where $SU(3)_C$ color symmetry would appear as a global or flavor symmetry due to the fact that the gluonic degrees of freedom are gapped by the Debye screening as well as the effects of Landau damping and thus they are integrated out. In this way, one can construct a holographic model of the CSC phase transition which describes the breaking of the global symmetries which consist of $SU(3)_C$ color symmetry, the usual $SU(N_f)$ flavor symmetry, and $U(1)_B$ symmetry.

In the strong coupling regime of QCD which is relevant to the core of neutron stars, a quark-gluon plasma contains both quarks of color-electric charge and composite scalars of color-magnetic charge. Their presence would thus generate the Debye screening masses $\left(m^2_E\right)_{ab}$ and $\left(m^2_M\right)_{ab}$ through the loop diagrams for the electric gluons $A^a_0$ and the magnetic gluons $A^a_i$, respectively. The electric and magnetic screening masses are determined as the static limit (i.e. $q_0=0$ and $\textbf{q}\rightarrow0$ with $(q_0,\textbf{q})\equiv q$ to be the four-momentum of gluons) of the temporal and spatial components of the gluon self-energy tensor $\Pi^{\mu\nu}_{ab}(q_0,\textbf{q})$, respectively, and given as follows \cite{Freedman1977,Kapusta1989,Bellac1996}
\begin{eqnarray}
\left(m^2_E\right)_{ab}&=&-\Pi^{00}_{ab}\left(0,\textbf{q}\rightarrow0\right)\sim g^2\left(T^2+\mu^2\right),\nonumber\\
\left(m^2_M\right)_{ab}&=&\frac{1}{2}\left(\delta^{ij}-\textbf{q}^i\textbf{q}^j\right)\Pi^{ij}_{ab}\left(0,\textbf{q}\rightarrow0\right)\sim g^2\left(T^2+\mu^2\right),
\end{eqnarray} 
where $g$, $T$, and $\mu$ are the strong coupling, the temperature of the quark-gluon plasma, and the chemical potential, respectively. In the regime of the CSC phase transition where the temperature is low enough and the quark matter density is high enough, the contribution of temperature is negligible compared to the contribution of quark matter density. Hence, the Debye screening masses of the electric and magnetic gluons are in order of the $g\mu$ scale. In the intermediate density regime of QCD, the strong coupling is $g\sim4\pi$ which is large, thus the superconductivity gap scale which is estimated as tens of MeV is much lower than the screening scale which is the hundreds of MeV. Hence, the electric and magnetic gluons are gapped or are Debye screened in the CSC phase which is much below the screening scale.

It should be noted that the QCD plasma at the weak coupling regime does not contain the composite scalars of color-magnetic charge \cite{Freedman1977} and since the Debye screening mass for the magnetic gluons can not be generated. This means that the magnetic gluons would lead to the dominant interaction in the momentum regime below the screening scale $g\mu$. However, the magnetic gluons can be gapped due to the effects of the Landau damping \cite{Son1999} where their self-energy behaves as $\Pi\sim g^2\mu^2|q_0|/|q|$ with the gap of electric gluons to be the source for the mass of the magnetic gluons.

In summary, all gluonic degrees of freedom which are color-electric and color-magnetic are screened at the momentum regime below the screening scale $g\mu$. Due to the large separation between the screening scale and the chemical potential scale, the gluons can be integrated out, and as a result, quarks exist as the sole degrees of freedom below the screening scale. On the other hand, in this momentum regime $SU(3)_C$ color symmetry can be realized as a global or a flavor symmetry.

\section{\label{GDM} Gravitational dual model}

Before introducing a gravitational dual model which is consistent with the CSC phase transition in the boundary field theory, let us briefly discuss the CSC condensation pattern which shall be considered in this work. A quark Cooper pair or diquark operator is formed from two quarks each of which carries the spin, color, and flavor degrees of freedom. For the formation of scalar spin-$0$ condensation, the combination of spins in the diquark operator would be antisymmetric. As mentioned in the introduction, in order to form the diquark operator we need to have the attractive interaction which appears in the color antitriplet channel. This means that the flavor combination in the diquark operator must be antisymmetric due to the Pauli principle which requires the wave function of the diquark operator to be antisymmetric under the exchange of two quarks. In the present work, we are interested in the presence of two lightest quark flavors (up and down quarks) participating in the Cooper pairing, whereas the strange quark is taken to be infinitely massive and hence it is absent. The corresponding CSC phase is well-known as the two-flavor CSC phase \cite{Alford2008}. In this way, the expectation value $\Delta$ of the diquark operator is given as
\begin{eqnarray}
\Delta=\left\langle q^TC\gamma_5\tau_2\lambda_2q\right\rangle,
\end{eqnarray}
which is antisymmetric in terms of the Dirac, color, and flavor indices that have been dropped. Here, $C\equiv i\gamma^2\gamma^0$, $\tau_2$, and $\lambda_2$ are the charge conjugation operator, the Pauli matrix, and the Gell-Mann matrix which act in the Dirac, flavor, and color spaces, respectively. In addition, the condensation of the diquark operator is triggered by the quark chemical potential or quark number density associated with the global $U(1)_B$ symmetry. Therefore, in the holographic model, we need to introduce a complex scalar field that is dual to $\Delta$ and carries a charge corresponding to a local $U(1)$ symmetry dual to the global $U(1)_B$ symmetry.

The gravitational dual model for the CSC phase transition is Einstein gravity coupled minimally to a power-law Maxwell field and a complex scalar field in the asymptotically six-dimensional AdS spacetime. The action of the system is given as follows
\begin{equation}
S_{\text{bulk}}=\frac{1}{2\kappa^2_6}\int
d^6x\sqrt{-g}\left[R-2\Lambda+\beta\left(-F_{\mu\nu}F^{\mu\nu}\right)^s-|(\nabla-iqA)\psi|^2-m^2|\psi|^2\right],\label{EGB-ED-adS}
\end{equation}
where $\Lambda=-\frac{10}{l^2}$ is the cosmological constant with $l$ being the asymptotic AdS radius, $\beta$ is a constant which is considered to be $1/4$ in the present work without loss of generality, $s$ is the power parameter characterizing for the power-law Maxwell field and in particular when $s=1$ the power-law Maxwell electrodynamics would reduce to the usual Maxwell case, and $\psi$ is the complex scalar field which has the mass $m$ and carries the charge $q$ under the gauge symmetry $U(1)$. It is important to note here that the gauge symmetry $U(1)$ and the complex scalar field in the bulk are dual to the baryon symmetry $U(1)_B$ and the diquark operator in the boundary field theory, respectively. Therefore, the $U(1)$ charge of the complex scalar field is expressed in terms of the color number of quarks as $q=2/N_c$. 

It is important to note that the four-dimensional gauge field theory possesses a confinement scale and hence the asymptotic behavior of spacetime geometry must manifest a corresponding scale. In the simplest setting \cite{Basu2011}, the gravity dual of the boundary field theory can be considered in six dimensions rather than five dimensions where one direction of spacetime is compactified on a circle whose size is identified as the confinement scale. In this way, the usually four-dimensional field theory can be obtained from the compactification of the five-dimensional field theory on a circle.

The equations of motion are found by varying the action (\ref{EGB-ED-adS}) with respect to the metric, vector, and scalar fields, given by
\begin{eqnarray}
R_{\mu\nu}-\frac{1}{2}g_{\mu\nu}R-\frac{10}{l^2}g_{\mu\nu}&=&T_{\mu\nu},\nonumber\\
\nabla_\mu\left[F^{\mu\nu}\left(-F_{\rho\sigma}F^{\rho\sigma}\right)^{s-1}\right]&=&\frac{iq}{s}\left[\psi^*(\nabla^\nu-iqA^\nu)\psi-\psi(\nabla^\nu+iqA^\nu)\psi^*\right],\nonumber\\
(\nabla_\mu-iqA_\mu)(\nabla^\mu-iqA^\mu)\psi&=&m^2\psi,\label{EOM}
\end{eqnarray}
where the energy-momentum tensor $T_{\mu\nu}$ reads
\begin{eqnarray}
T_{\mu\nu}&=&\frac{s}{2}\left(-F_{\rho\sigma}F^{\rho\sigma}\right)^{s-1}F_{\mu\lambda}{F_\nu}^\lambda+\frac{1}{2}\left[(\nabla_\nu-iqA_\nu)\psi(\nabla_\mu+iqA_\mu)\psi^*+\mu\leftrightarrow\nu\right]\nonumber\\
&&+\frac{1}{2}g_{\mu\nu}\left[\frac{1}{4}\left(-F_{\mu\nu}F^{\mu\nu}\right)^s-|(\nabla-iqA)\psi|^2-m^2|\psi|^2\right].
\end{eqnarray}
By solving these equations of motion in the spacetime geometries dual to the confinement and deconfinement phases, we shall find whether or not the CSC phase transition can appear in these phases. The CSC phase transition appears when the scalar field condenses where the $U(1)$ gauge symmetry is spontaneously broken. In the canonical ensemble, the condensation of the scalar field or the CSC phase transition is triggered by the chemical potential and for the chemical potential above a critical value $\mu_c$ the CSC phase transition would appear. In this work, we study the CSC phase transition near the critical chemical potential $\mu_c$ and since the condensate value of the scalar field is near zero. This suggests that the backreaction of the scalar field on the spacetime geometry can be ignored compared to the contribution of the vector field.

The spacetime geometry which is dual to the deconfinement phase is the high-temperature solution of Eq. (\ref{EOM}) known as the planar AdS black hole. Ansatz for the line element of the planar AdS black hole is given as follows 
\begin{eqnarray}
ds^2_{\text{BH}}&=&r^2\left(-f(r)dt^2+h_{ij}dx^idx^j+dy^2\right)+\frac{dr^2}{r^2f(r)},\label{BHa}
\end{eqnarray}
where $h_{ij}dx^idx^j=dx^2_1+dx^2_2+dx^2_3$ refers to the line element of the $3$-dimensional planar hypersurface and $y$ is the compactified coordinate with the circle radius $R_y$. Whereas, the vector and scalar fields are described by the following ansatz
\begin{eqnarray}
A_\mu dx^\mu=\phi(r)dt,\ \ \ \ \psi=\psi(r).\label{vec-scal-ans}
\end{eqnarray}
The equations of motion corresponding to the geometric configuration of the planar AdS black hole are found as
\begin{eqnarray}
rf'(r)+5f(r)-5+(2s-1)2^{s-4}\phi'(r)^{2s}&=&0,\label{r-f-Eq}\\
\phi''(r)+\frac{4}{(2s-1)r}\phi'(r)-\frac{q^2\psi^2(r)\phi'(r)^{2(1-s)}}{2^{s-2}s(2s-1)r^2f(r)}\phi(r)&=&0,\label{r-phi-Eq}\\
\psi''(r)+\left[\frac{f'(r)}{f(r)}+\frac{6}{r}\right]\psi'(r)+\frac{1}{r^2f(r)}\left[\frac{q^2\phi^2(r)}{r^2f(r)}-m^2\right]\psi(r)&=&0.\label{r-psi-Eq}
\end{eqnarray}
In order for the matter fields to behave regularly at the event horizon, we impose the following boundary condition
\begin{eqnarray}
\phi(r_+)=0,\ \ \ \ \psi(r_+)=r^2_+\frac{f'(r_+)\psi'(r_+)}{m^2}.\label{DCP-bc}
\end{eqnarray}
In the asymptotic region $r\rightarrow\infty$, the matter fields behave as
\begin{eqnarray}
\phi(r)&=&\mu-\frac{\bar{d}^{\frac{1}{2s-1}}}{r^{\frac{5-2s}{2s-1}}},\nonumber\\
\psi(r)&=&\frac{J_C}{r^{\Delta_-}}+\frac{C}{r^{\Delta_+}}.\label{phi-psi-asy-beh}
\end{eqnarray}
where $\mu$ and $\bar{d}$ are regarded as the baryon chemical potential and the baryon charge density of the boundary field theory, respectively. The overall coefficients $J_C$ and $C$ which both are normalizable modes are identified as the external source and the condensate value of the diquark operator, respectively. In order to guarantee that the $U(1)$ gauge symmetry is spontaneously broken, the external source must vanish and hence we impose a boundary condition as $J_C=0$. The conformal dimensions of $C$ and $J_C$ are
\begin{eqnarray}
\Delta_\pm=\frac{1}{2}\left(5\pm\sqrt{25+4m^2}\right),
\end{eqnarray}
from which we see that the squared mass of the scalar field must satisfy the Breitenlohner-Freedman (BF) bound \cite{Freedman1982,Breitenlohner1982} as
\begin{equation}
m^2\geq-\frac{25}{4}.
\end{equation}
Because $C$ is identified as the condensate value of the diquark operator, the conformal dimension of $C$ should be $\Delta_+=2\times(d-2)/2$ leading to $\Delta_+=4$ for the six-dimensional case. This means that the squared mass of the scalar field is $m^2=-4$. In addition, we note that, from the asymptotic expression of the scalar potential $\phi(r)$, in order for $\phi(r)$ to be finite as $r\rightarrow\infty$, the power parameter $s$ must satisfy the following condition
\begin{eqnarray}
\frac{1}{2}<s<\frac{5}{2}.
\end{eqnarray}

In the limit that the chemical potential approaches the critical value $\mu_c$, we obtain
\begin{eqnarray}
f(r)&=&1-\left(\frac{r_+}{r}\right)^5-\frac{2^{s-4}(2s-1)^2}{5-2s}\left[\frac{(5-2s)\mu}{(2s-1)r_+}\right]^{2s}\left(\frac{r_+}{r}\right)^5\left[1-\left(\frac{r_+}{r}\right)^{\frac{5-2s}{2s-1}}\right],\label{fr-CAdSBH}\nonumber\\
\phi(r)&=&\mu\left[1-\left(\frac{r_+}{r}\right)^{\frac{5-2s}{2s-1}}\right].
\end{eqnarray}
The Hawking temperature of the planar AdS black hole thus is
\begin{eqnarray}
T=\frac{r_+}{4\pi}\left\{5-2^{s-4}(2s-1)\left[\frac{(5-2s)\mu}{(2s-1)r_+}\right]^{2s}\right\}.
\end{eqnarray}
The non-negative condition of the Hawking temperature leads to the condition for the ratio of the chemical potential to the event horizon radius as
\begin{eqnarray}
0\leq\frac{\mu}{r_+}\leq\frac{2s-1}{5-2s}\left[\frac{80}{2^s(2s-1)}\right]^{\frac{1}{2s}}.\label{red-cp-cond}
\end{eqnarray}

The spacetime geometry which is dual to the confinement phase is the low-temperature solution of Eq. (\ref{EOM}) known as the AdS soliton solution\footnote{The low-temperature solution of AdS soliton was also confirmed in the extensions of Einstein gravity, for example, in EGB gravity \cite{Cai-Kim2007}.} \cite{Horowitz1998} described by the following ansatz
\begin{eqnarray}
ds^2_{\text{Sol.}}=r^2\left(-dt^2+h_{ij}dx^idx^j+f(r)dy^2\right)+\frac{dr^2}{r^2f(r)},\label{AdSs}
\end{eqnarray}
where the function $f(r)$ vanishes at the tip $r=r_0$ of the AdS soliton where the canonical singularity is removed by requiring the periodicity for the coordinate $y$. The equations of motion corresponding to this geometric configuration are
\begin{eqnarray}
rf'(r)+5f(r)-5+(2s-1)2^{s-4}f(r)^s\phi'(r)^{2s}&=&0,\\
\phi''(r)+\frac{1}{2s-1}\left[s\frac{f'(r)}{f(r)}+\frac{4}{r}\right]\phi'(r)-\frac{q^2\psi^2(r)\phi'(r)^{2(1-s)}}{2^{s-2}s(2s-1)r^2f(r)^s}\phi(r)&=&0,\\
\psi''(r)+\left[\frac{f'(r)}{f(r)}+\frac{6}{r}\right]\psi'(r)+\frac{1}{r^2f(r)}\left[\frac{q^2\phi^2(r)}{r^2}-m^2\right]\psi(r)&=&0.\label{AdSsol-psi-eq}
\end{eqnarray}
The boundary condition for the matter fields is
\begin{eqnarray}
\phi'(r_0)&=&\left[\frac{q^2\psi^2(r_0)}{2^{s-2}s^2r^2_0f(r_0)^{s-1}f'(r_0)}\phi(r_0)\right]^{\frac{1}{2s-1}},\\
\psi'(r_0)&=&-\frac{1}{r^2_0f'(r_0)}\left[\frac{q^2\phi^2(r_0)}{r^2_0}-m^2\right]\psi(r_0).
\end{eqnarray}
The asymptotic behavior of the matter fields is the same as in Eq. (\ref{phi-psi-asy-beh}). Near the critical chemical potential where the value of the scalar field approaches zero, the AdS soliton solution is determined by the line element (\ref{AdSs}) with the function $f(r)$ given as
\begin{eqnarray}
f(r)=1-\frac{r^5_0}{r^5},\ \ \ \ r_0=\frac{2}{5R_y},\label{AdSsol-fr}
\end{eqnarray}
and the potential of the $U(1)$ gauge field reads
\begin{eqnarray}
\phi(r)=\mu=\text{constant}.\label{AdSs-phir}
\end{eqnarray}

\section{\label{CSCPT} CSC phase transition}

In this section, we will study the CSC phase transition of the realistic YM theory in the background configurations dual to the confinement and deconfinement phases by examining the breaking of the BF bound and solving numerically the equations of motion near the critical chemical potential. We point out that the CSC phase transition with $N_c\geq2$ in the deconfinement phase appears above the critical chemical potential corresponding to the condensation of the scalar field, which is not found in the case of the usual Maxwell electrodynamics if the power parameter $s$ is sufficiently smaller than one. Furthermore, the value of the power parameter $s$ is not arbitrary below one (but above $1/2$). This is because when $s$ is sufficiently far away from one it leads to the condensation of the scalar field in the confinement phase, i.e. the occurrence of the CSC state in the confinement phase, which suggests the breakdown of the holographic description.

\subsection{In the deconfinement phase}

\begin{figure}[t]
 \centering
\begin{tabular}{cc}
\includegraphics[width=0.45 \textwidth]{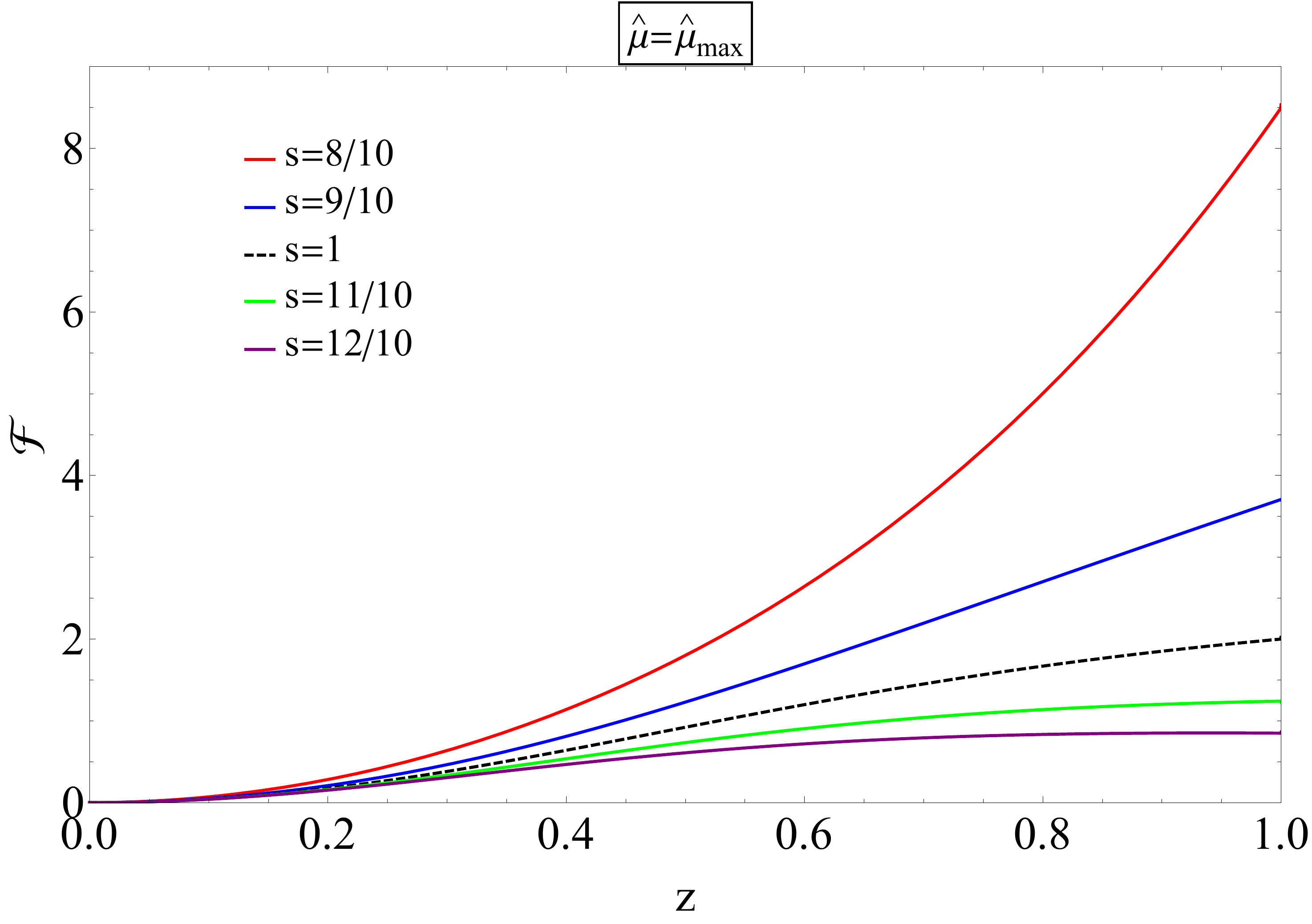}
\hspace*{0.05\textwidth}
\includegraphics[width=0.45 \textwidth]{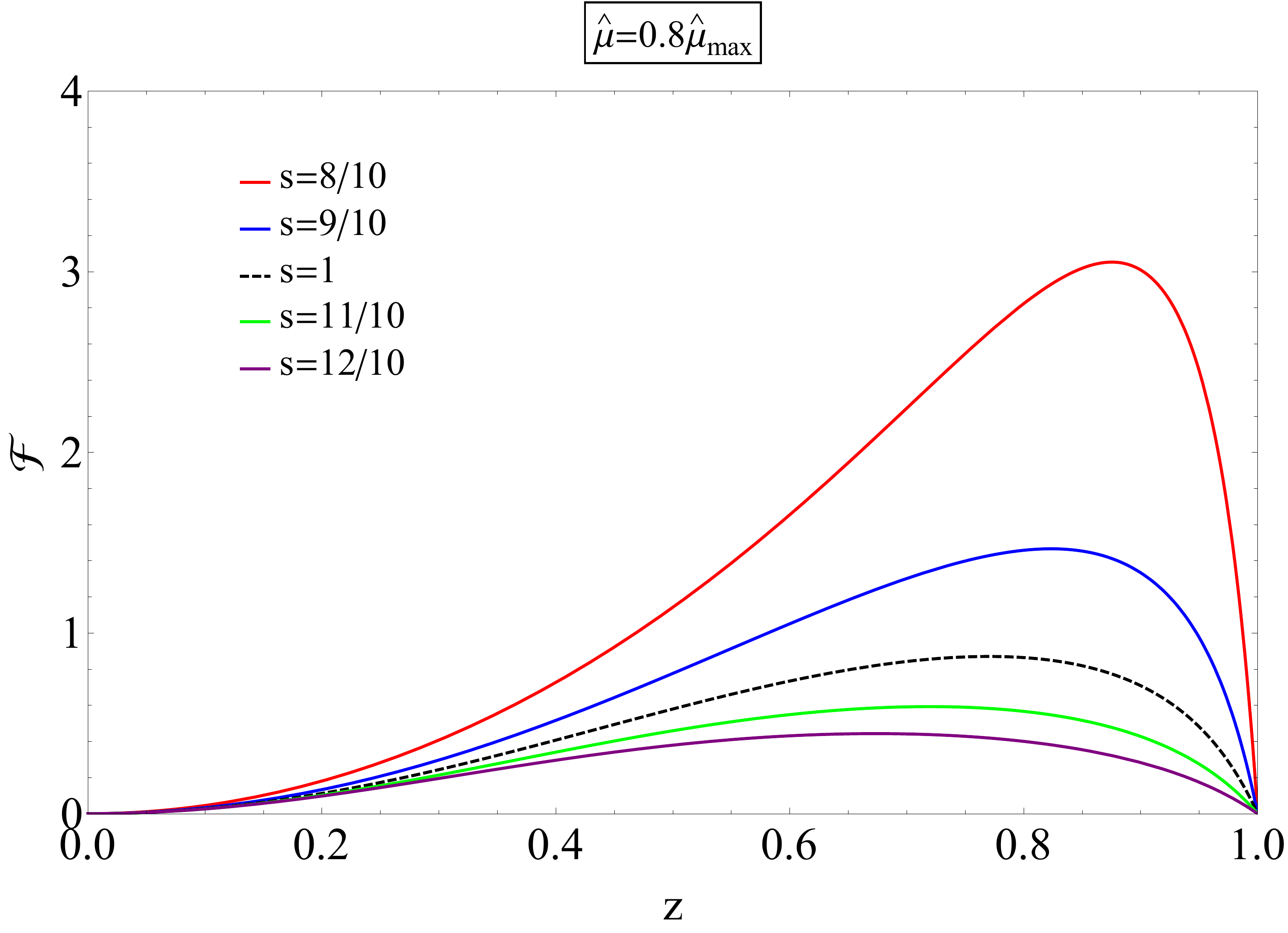}\\
\includegraphics[width=0.45 \textwidth]{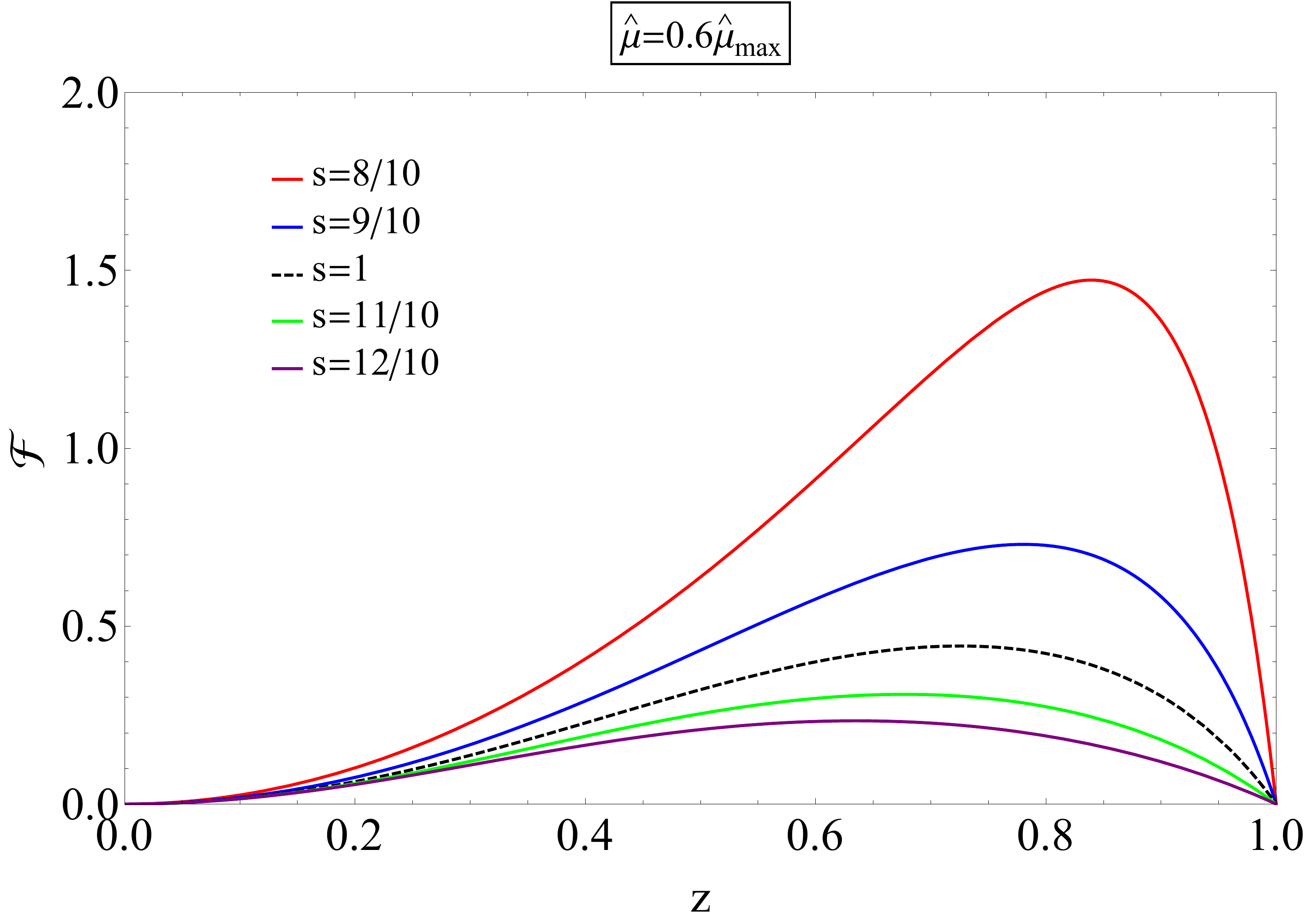}
\hspace*{0.05\textwidth}
\includegraphics[width=0.45 \textwidth]{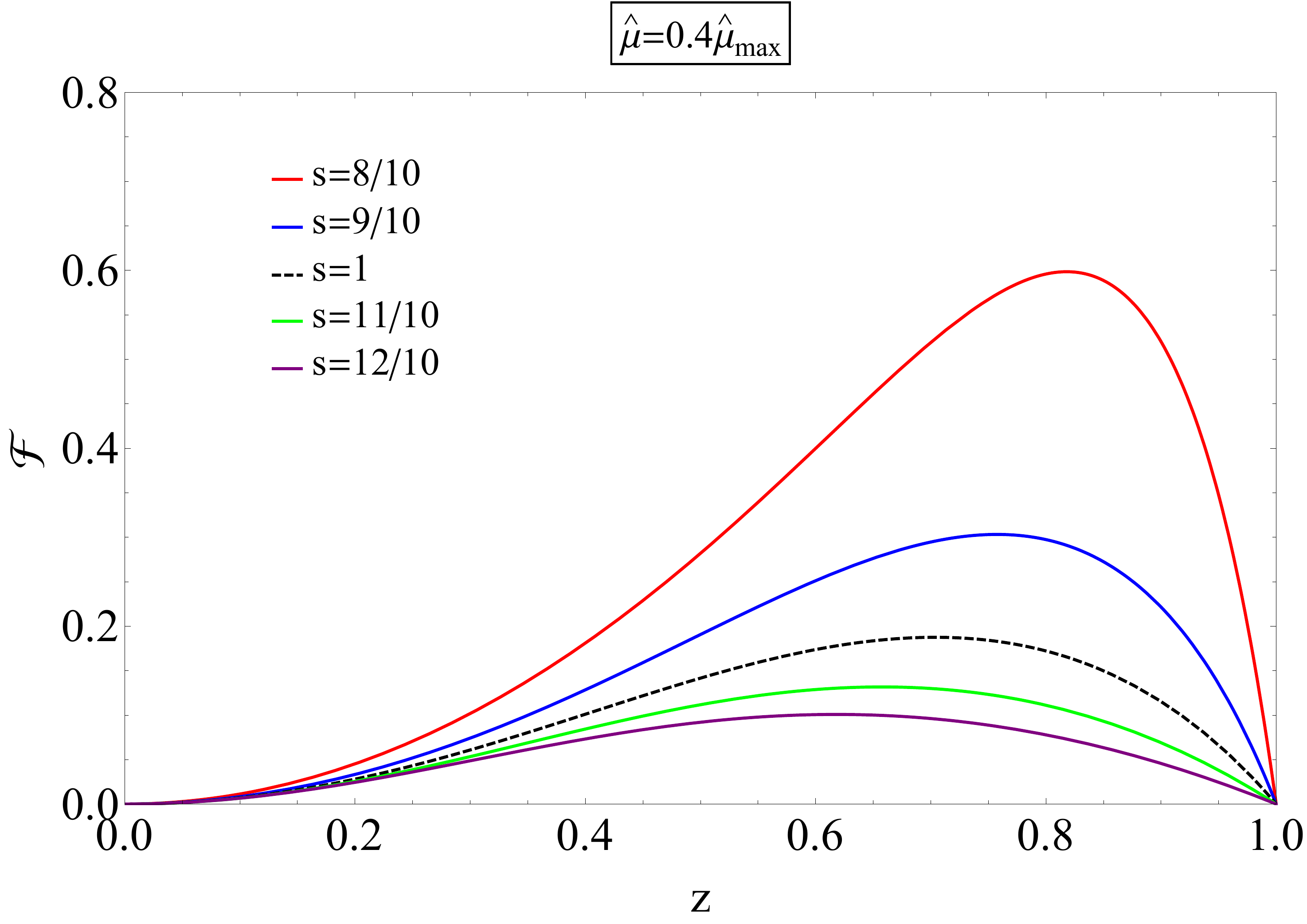}
\end{tabular}
  \caption{Plots describing the behavior of $\mathcal{F}(z,\hat{\mu},\alpha)$ in terms of $z$, $\hat{\mu}$ and $\alpha$, where $\hat{\mu}_{\text{max}}$ is given in Eq. (\ref{mumax}).}\label{F-beh}
\end{figure}

\begin{figure}[t]
 \centering
\begin{tabular}{cc}
\includegraphics[width=0.6 \textwidth]{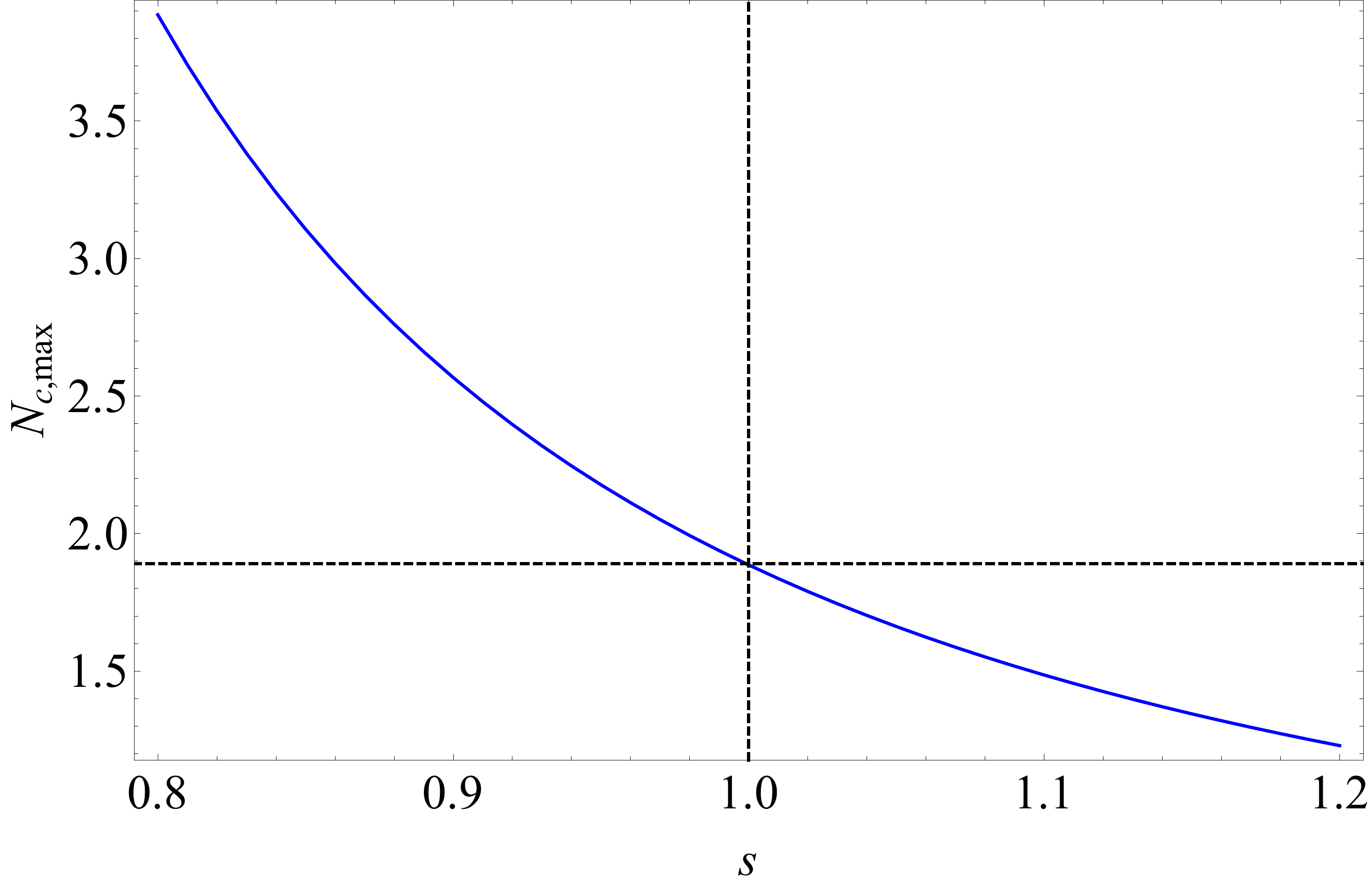}
\end{tabular}
 \caption{The upper bound for the color number of quarks is plotted in the power parameter $s$. The dashed black lines refer to the case of the usual Maxwell electrodynamics.}\label{Nc-upbound}
\end{figure}

In order to see that the CSC phase transition for $N_c\geq2$ can appear in the holographic model with the power-law Maxwell field for the appropriate value of the power parameter $s$, first let us study the condition which breaks the BF bound and thus makes the condensation of the scalar field appearing. The condition of the BF bound breaking is given as
\begin{eqnarray}
m^2_{\text{eff}}<-\frac{25}{4} \ \ \ \ \Rightarrow \ \ \ \ \frac{9}{4}<\frac{q^2\phi^2(r)}{r^2f(r)}\equiv q^2\mathcal{F}(z,\hat{\mu},s),\label{BFcond}
\end{eqnarray}
where $m^2_{\text{eff}}=m^2-\frac{q^2\phi^2(r)}{r^2f(r)}$ is the effective squared mass of the scalar field and
\begin{eqnarray}
\mathcal{F}(z,\hat{\mu},s)=\frac{z^2\left(1-z^{\frac{5-2s}{2s-1}}\right)^2\hat{\mu}^2}{1-z^5-\frac{2^{s-4}(2s-1)^2}{5-2s}\left[\frac{(5-2s)\hat{\mu}}{(2s-1)}\right]^{2s}z^5\left(1-z^{\frac{5-2s}{2s-1}}\right)},
\end{eqnarray}
with $z\equiv r_+/r$ and $\hat{\mu}\equiv\mu/r_+$. We show the behavior of the function $\mathcal{F}(z,\hat{\mu},s)$ as a function of $z$ for various values of $\hat{\mu}$ and $s$ in Fig. \ref{F-beh}. From this figure, we realize that $\mathcal{F}(z,\hat{\mu},s)$ grows with the increasing of the scaled chemical potential $\hat{\mu}$ and the decreasing of the power parameter $s$.
Furthermore, we find that the function $\mathcal{F}(z,\hat{\mu},s)$ gets the maximum value when $z\rightarrow1$ and at
\begin{eqnarray}
\hat{\mu}=\frac{2s-1}{5-2s}\left[\frac{80}{2^s(2s-1)}\right]^{\frac{1}{2s}}\equiv\hat{\mu}_{\text{max}}.\label{mumax}
\end{eqnarray}
From this result along with Eq. (\ref{BFcond}), we obtain an upper bound for the color number, which depends on the power parameter $s$, as
\begin{eqnarray}
N_c<\frac{4}{3}\sqrt{\mathcal{F}(z\rightarrow1,\hat{\mu}_{\text{max}},s)}\equiv N_{c,\text{max}}.
\end{eqnarray}
The upper bound $N_{c,\text{max}}$ for the color number of quarks as a function of the power parameter $s$ is depicted in Fig. \ref{Nc-upbound}.
We see here that $N_{c,\text{max}}$ increases when decreasing $s$, and $N_{c,\text{max}}$ in the power-law Maxwell electrodynamics with $s<1(>1)$ is larger(smaller) than that ($N_c<4\sqrt{2}/3\simeq1.89$ \cite{Ghoroku2019}) in the usual Maxwell electrodynamics. In particular, we find that the upper bound for the color number of quarks is much larger than $2$ for the appropriate value of the power parameter $s$ belonging to the region $1/2<s<1$. This clearly suggests that the power-law Maxwell electrodynamics with $s<1$ can support the CSC phase transition for the realistic YM theory which has $N_c\geq2$. 

In order to determine the critical chemical potential, the critical temperature, and the slope of the critical line $T_c=T_c(\mu_c)$ associated with the CSC phase transition in the deconfinement phase, we solve numerically Eqs. (\ref{r-f-Eq}), (\ref{r-phi-Eq}), and (\ref{r-psi-Eq}) by using the shooting method. The numerical results are given in Tables \ref{Nc2} and \ref{Nc3} for $N_c=2$ and $N_c=3$, respectively.
\begin{table}[!htp]
\centering
\begin{tabular}{c|c|c|c}
  \hline
  \hline
  $s$ & $\mu_c/r_+$ & $T_c/r_+$ & $T_c/\mu_c$ \\
  \hline
  $\ \ \ \ \ \ 4/5 \ \ \ \ \ \ $ & $\ \ \ \ \ \ 2.5413 \ \ \ \ \ \ $ & $\ \ \ \ \ \ 0.0271 \ \ \ \ \ \ $ & $\ \ \ \ \ \ 0.0107 \ \ \ \ \ \ $\\
  \hline
  $\ \ \ \ \ \ 3/4 \ \ \ \ \ \ $ & $\ \ \ \ \ \ 2.6192 \ \ \ \ \ \ $ & $\ \ \ \ \ \ 0.0696 \ \ \ \ \ \ $ & $\ \ \ \ \ \ 0.0266 \ \ \ \ \ \ $\\
  \hline
  $\ \ \ \ \ \ 5/7 \ \ \ \ \ \ $ & $\ \ \ \ \ \ 2.6299 \ \ \ \ \ \ $ & $\ \ \ \ \ \ 0.1101 \ \ \ \ \ \ $ & $\ \ \ \ \ \ 0.0418 \ \ \ \ \ \ $\\
  \hline
  \hline
\end{tabular}
\caption{The numerical values for the scaled critical chemical potential and temperature, and the slope of the critical line $T_c=T_c(\mu_c)$ with various values of $s$ at $N_c=2$.} \label{Nc2}
\end{table}
\begin{table}[!htp]
\centering
\begin{tabular}{c|c|c|c}
  \hline
  \hline
  $s$ & $\mu_c/r_+$ & $T_c/r_+$ & $T_c/\mu_c$ \\
  \hline
  $\ \ \ \ \ \ 2/3 \ \ \ \ \ \ $ & $\ \ \ \ \ \ 3.4058 \ \ \ \ \ \ $ & $\ \ \ \ \ \ 0.0680 \ \ \ \ \ \ $ & $\ \ \ \ \ \ 0.0200 \ \ \ \ \ \ $\\
  \hline
  $\ \ \ \ \ \ 5/8 \ \ \ \ \ \ $ & $\ \ \ \ \ \ 3.4375 \ \ \ \ \ \ $ & $\ \ \ \ \ \ 0.1329 \ \ \ \ \ \ $ & $\ \ \ \ \ \ 0.0387 \ \ \ \ \ \ $\\
  \hline
  $\ \ \ \ \ \ 3/5 \ \ \ \ \ \ $ & $\ \ \ \ \ \ 3.3839 \ \ \ \ \ \ $ & $\ \ \ \ \ \ 0.1750 \ \ \ \ \ \ $ & $\ \ \ \ \ \ 0.0517 \ \ \ \ \ \ $\\
  \hline
  \hline
\end{tabular}
\caption{The numerical values for the ratios $\mu_c/r_+$, $T_c/r_+$, and $T_c/\mu_c$ with various values of $s$ at $N_c=3$.} \label{Nc3}
\end{table}
These tables suggest that there exist the proper values of the power parameter $s$ which are sufficiently low and below one (but still belonging to the physical region) for which the CSC phase transition with $N_c\geq2$ can appear. In addition, we observe that for the event horizon radius $r_+$ kept fixed the critical temperature grows when the power parameter $s$ decreases. This means that the condensation of the scalar field gets easier to form in the power-law Maxwell electrodynamics with the appropriately low power parameter.

\begin{figure}[t]
 \centering
\begin{tabular}{cc}
\includegraphics[width=0.6 \textwidth]{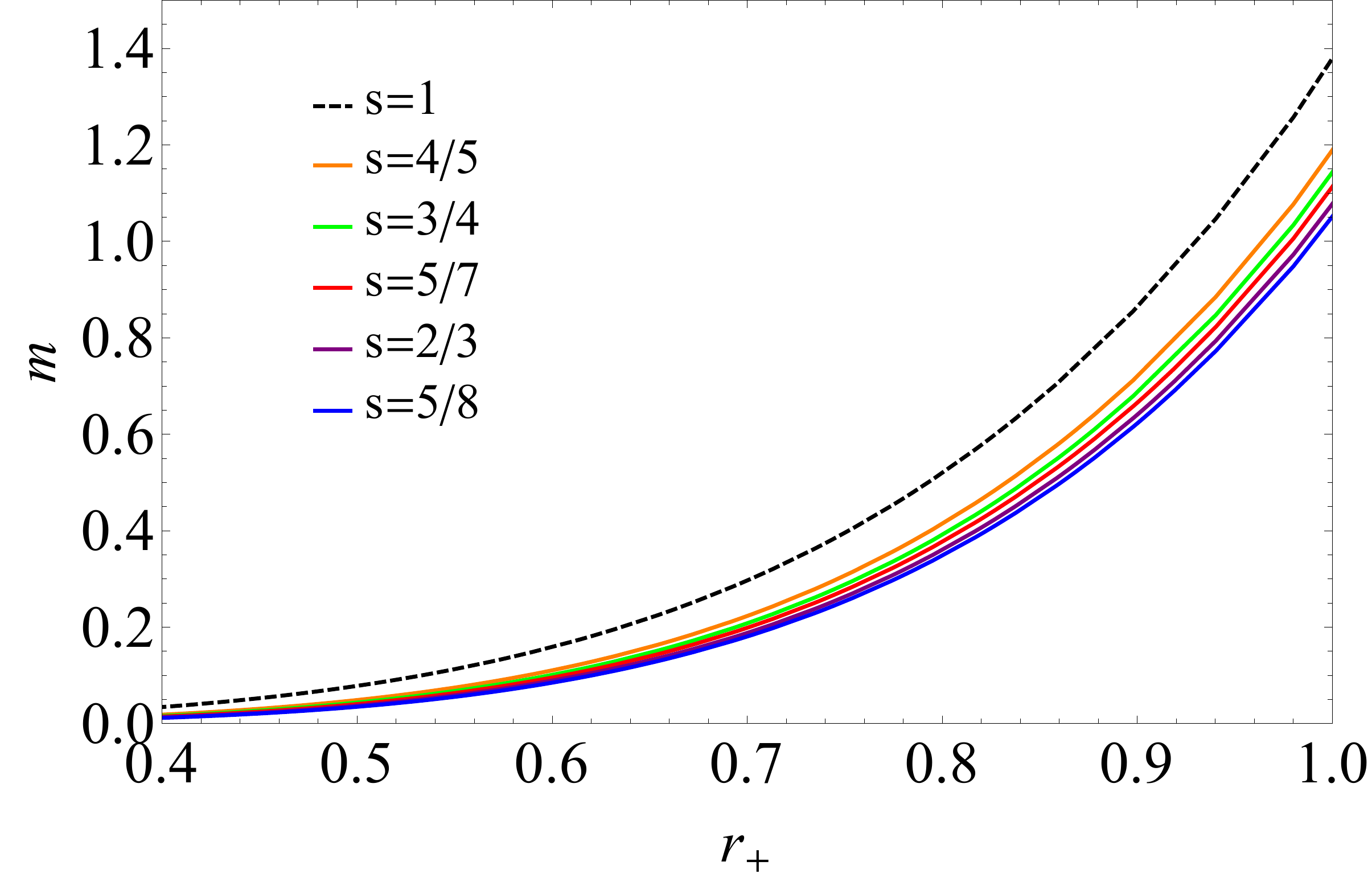}
\end{tabular}
  \caption{The mass $m$ of the planar AdS black hole which can be inferred from the terms proportional to $1/r^5$ in Eq. (\ref{fr-CAdSBH}) in terms of the event horizon radius $r_+$ for various values of the power parameter $s$.}\label{BHmass-ehrad}
\end{figure}

Let us interpret why the power-law Maxwell electrodynamics with the appropriately low power parameter can lead to the occurrence of the CSC phase transition for $N_c\geq2$. Recall that the CSC phase of the realistic YM theory which corresponds to $N_c\geq2$ is not found in both the deconfinement and the confinement phases with respect to the holographic model which consists of the Einstein-Maxwell system coupled to the complex scalar field \cite{Ghoroku2019}. When the color number of quarks increases the charge of the complex scalar field would reduce. As a result, the electrostatic repulsion becomes weak and since it would not be strong enough to overcome the gravitational attraction for the scalar hair formed. However, when we consider the power-law Maxwell electrodynamics instead of the usual Maxwell electrodynamics, the gravitational attraction would become weak with the decrease of the power parameter $s$. This can be seen in Fig. \ref{BHmass-ehrad} where we observe that with the same mass the event horizon of the planar AdS black hole grows with the decreasing of the power parameter $s$.
On the other hand, the gravitational attraction around the event horizon is sufficiently weak with the properly low power parameter, by which the electrostatic repulsion can overcome the gravitational attraction to allow the formation of the scalar hair corresponding to the CSC phase transition even with the low charge of the complex scalar field or $N_c\geq2$.

\subsection{In the confinement phase}
We arrive at the investigation of whether or not the CSC state or the condensation of the scalar field appears in the confinement phase. It is easy to find a necessary condition for the appearance of the CSC phase transition in this phase as $q\mu/r_0>1.5$, which corresponds to the breaking of the BF bound for the effective squared mass $m^2_{\text{eff}}=m^2-q^2\phi^2/r^2<-25/4$. Of course, in order to determine a sufficient condition for the appearance of the CSC phase transition in the confinement phase, we need to solve Eq. (\ref{AdSsol-psi-eq}) in the background configuration of the AdS soliton determined by Eqs. (\ref{AdSs}), (\ref{AdSsol-fr}), and (\ref{AdSs-phir}). By solving numerically Eq. (\ref{AdSsol-psi-eq}), we find the condition of the chemical potential for the appearance of the CSC phase transition in the confinement phase as $\mu>1.505N_c$. However, in order to confirm that the CSC phase transition can appear in the confinement phase when the chemical potential is above the critical value $1.505N_c$, we need to check whether or not this critical value belongs to the region of the confinement phase. By using the results in the next section, we show the maximal chemical potential $\mu_{\text{max}}$ which the confinement phase can reach as well as the critical chemical potential $\mu_c$ for the occurrence of the CSC state in the confinement phase for various values of the power parameter $s$ at $N_c=2$ and $N_c=3$ in Table \ref{CP-CSC}.
\begin{table}[!htp]
\centering
\begin{tabular}{c|c|c||c|c|c}
  \hline
  \hline
  \multicolumn{3}{c||}{$N_c=2$} & \multicolumn{3}{c}{$N_c=3$}\\
  \hline
  $s$ & $\mu_{\text{max}}$ & $\mu_c$ & $s$ & $\mu_{\text{max}}$ & $\mu_c$\\
  \hline
  $\ \ \ \ 4/5\ \ \ \ $ & $\ \ \ \ 2.337\ \ \ \ $ & $\ \ \ \ 3.01\ \ \ \ $ & $\ \ \ \ 2/3\ \ \ \ $ & $\ \ \ \ 3.630\ \ \ \ $ &  $\ \ \ \ 4.515\ \ \ \ $\\
  \hline
  $\ \ \ \ 3/4\ \ \ \ $ & $\ \ \ \ 2.672\ \ \ \ $ & $\ \ \ \ 3.01\ \ \ \ $ & $\ \ \ \ 5/8\ \ \ \ $ & $\ \ \ \ 4.492\ \ \ \ $ &  $\ \ \ \ 4.515\ \ \ \ $\\
  \hline
  $\ \ \ \ 5/7\ \ \ \ $ & $\ \ \ \ 3.007\ \ \ \ $ & $\ \ \ \ 3.01\ \ \ \ $ & $\ \ \ \ 3/5\ \ \ \ $ & $\ \ \ \ 5.230\ \ \ \ $ &  $\ \ \ \ 4.515\ \ \ \ $\\
  \hline
  \hline
\end{tabular}
\caption{The numerical values for the maximal chemical potential $\mu_{\text{max}}$ above which the confinement phase cannot appear and the critical chemical potential $\mu_c$ above which the CSC phase transition happens for various values of the power parameter $s$ at $N_c=2$ and $N_c=3$.}\label{CP-CSC}
\end{table}
Note that, the maximal chemical potential $\mu_{\text{max}}$ can be found from Eq. (\ref{peq-DCCnound}) with $T(r_+)=0$. Clearly, this table indicates that the critical chemical potential $\mu_c=3.01$ and $\mu_c=4.515$ corresponding to $N_c=2$ and $N_c=3$, respectively, are outside the confinement phase for most values of the power parameter $s$ under consideration. This means that the CSC phase transition cannot occur in the confinement phase except in the case of $s=3/5$ corresponding to $N_c=3$. Also, we observe that the maximal chemical potential $\mu_{\text{max}}$ grows with the decreasing of the power parameter $s$ and as a result $\mu_{\text{max}}$ would be larger than the critical chemical potential $\mu_c$ at the sufficiently low values of the power parameter $s$. This suggests that the CSC state can exist even in the confinement phase which would not be compatible with the nonzero VEV of the color nonsinglet operator. On the other hand, the present holographic model with sufficiently low values of the power parameter $s$ does not provide a suitable holographic model for the CSC phase transition of the realistic YM theory.
\section{\label{PhaseD} Phase diagram}

In this section, we will obtain the phase diagram of the present holographic model in the plane of the temperature $T$ and the chemical potential $\mu$, which shows the information about the region of $T$ and $\mu$ where the thermodynamically distinct phases (such as the deconfinement, confinement, and CSC phases which we are interested in this work) appear. In addition, the phase diagram shows the location of the phase boundaries or the critical curves and when there is a phase transition between the different phases.

First, we determine the phase structure which describes the confinement and deconfinement phases, which corresponds to the vanishing of the scalar field. In order to do this, we need to compute the free energy of the AdS soliton configuration and the planar AdS black hole configuration from the Euclidean on-shell action of the system and then compare them to find which one is the preferred configuration. By using the general result obtained in \cite{Miskovic2011}, one can write the free energy of Einstein gravity coupled minimally to the power-law Maxwell field in the situation without the condensation of the scalar field as 
\begin{eqnarray}
S_E&=&\left[r^4\left(r^2f\right)'\Big|^\infty_{r_+}-r^4f^2\left(r^2f\right)'\Big|^\infty-s2^{s-1}\int^{\infty}_{r_+}drr^4\phi'(r)^{2s}\right]\frac{4\pi}{5r_0}\frac{V_3}{T}\nonumber\\
&=&\left[r^4\left(r^2f\right)'\Big|^\infty_{r_+}-r^4f^2\left(r^2f\right)'\Big|^\infty-s2^{s-1}r^4\phi\phi'^{2s-1}\Big|^\infty_{r_+}\right]\frac{4\pi}{5r_0}\frac{V_3}{T},\label{OSEA}
\end{eqnarray}
where $V_3=\int dx_1dx_2dx_3$ and in the second line we have used the equation of motion for $\phi(r)$ as follows
\begin{eqnarray}
\frac{d}{dr}\left[r^4\phi'(r)^{2s-1}\right]=0.
\end{eqnarray}
Then, we obtain explicitly the free energy for the planar AdS black hole configuration and the AdS soliton configuration as
\begin{eqnarray}
\Omega_{\text{BH}}&=&-r^5_+\left[1+\frac{2^{s-4}(2s-1)^2}{(5-2s)}\left(\frac{(5-2s)\mu}{(2s-1)r_+}\right)^{2s}\right]\frac{4\pi}{5r_0}V_3,\nonumber\\
\Omega_{\text{Sol.}}&=&-r^5_0\frac{4\pi}{5r_0}V_3.\label{FE-BH-AdSS}
\end{eqnarray}
The free energy of the planar AdS black hole and AdS soliton is the same, i.e. $\Omega_{\text{BH}}=\Omega_{\text{Sol.}}$, along the critical curve which defines the phase boundary between the confinement and deconfinement phases. From this relation, we find the parameter equation in terms of the event horizon radius $r_+$, which determines this critical curve in the $\mu-T$ plane as
\begin{eqnarray}
\mu(r_+)&=&\left[\frac{2^{4-s}(1-r^5_+)}{(2s-1)r^{5-2s}_+}\right]^{1/2s}\left(\frac{5-2s}{2s-1}\right)^{(1-2s)/2s},\nonumber\\
T(r_+)&=&\frac{r_+}{4\pi}\left[5-\frac{(5-2s)}{(2s-1)}\frac{(1-r^5_+)}{r^5_+}\right],\label{peq-DCCnound}
\end{eqnarray}
where we have set $r_0$ to be unity without loss of generality. The confinement phase corresponds to the region in the $\mu-T$ plane where the AdS soliton has the free energy lower than that of the planar AdS black hole and since the geometric configuration of the AdS soliton is thermodynamically favored. On the contrary, the region where the free energy of the planar AdS black hole is lower than that of the AdS soliton represents the deconfinement phase. 

\begin{figure}[t]
 \centering
\begin{tabular}{cc}
\includegraphics[width=0.45 \textwidth]{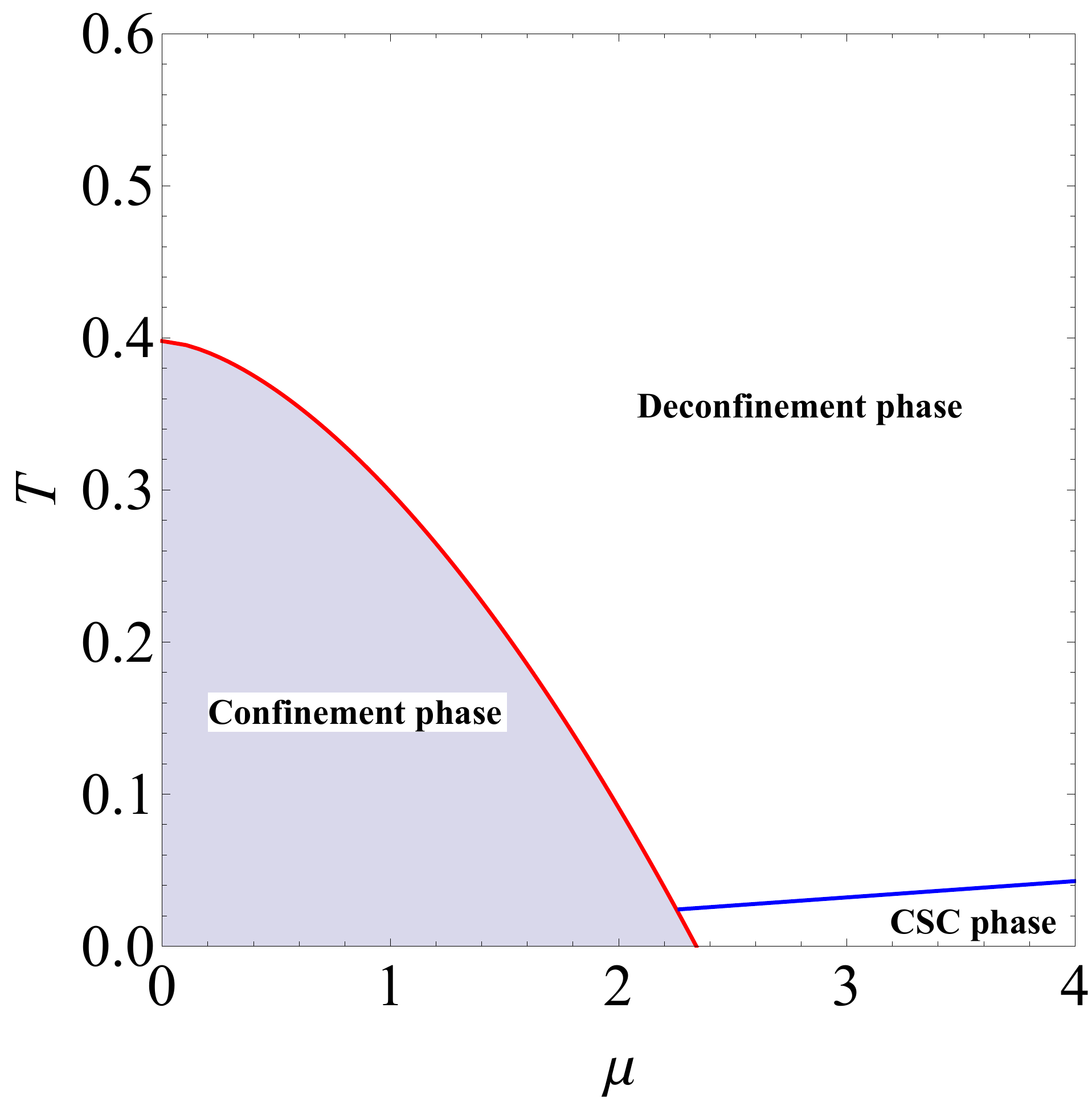}
\hspace*{0.05\textwidth}
\includegraphics[width=0.45 \textwidth]{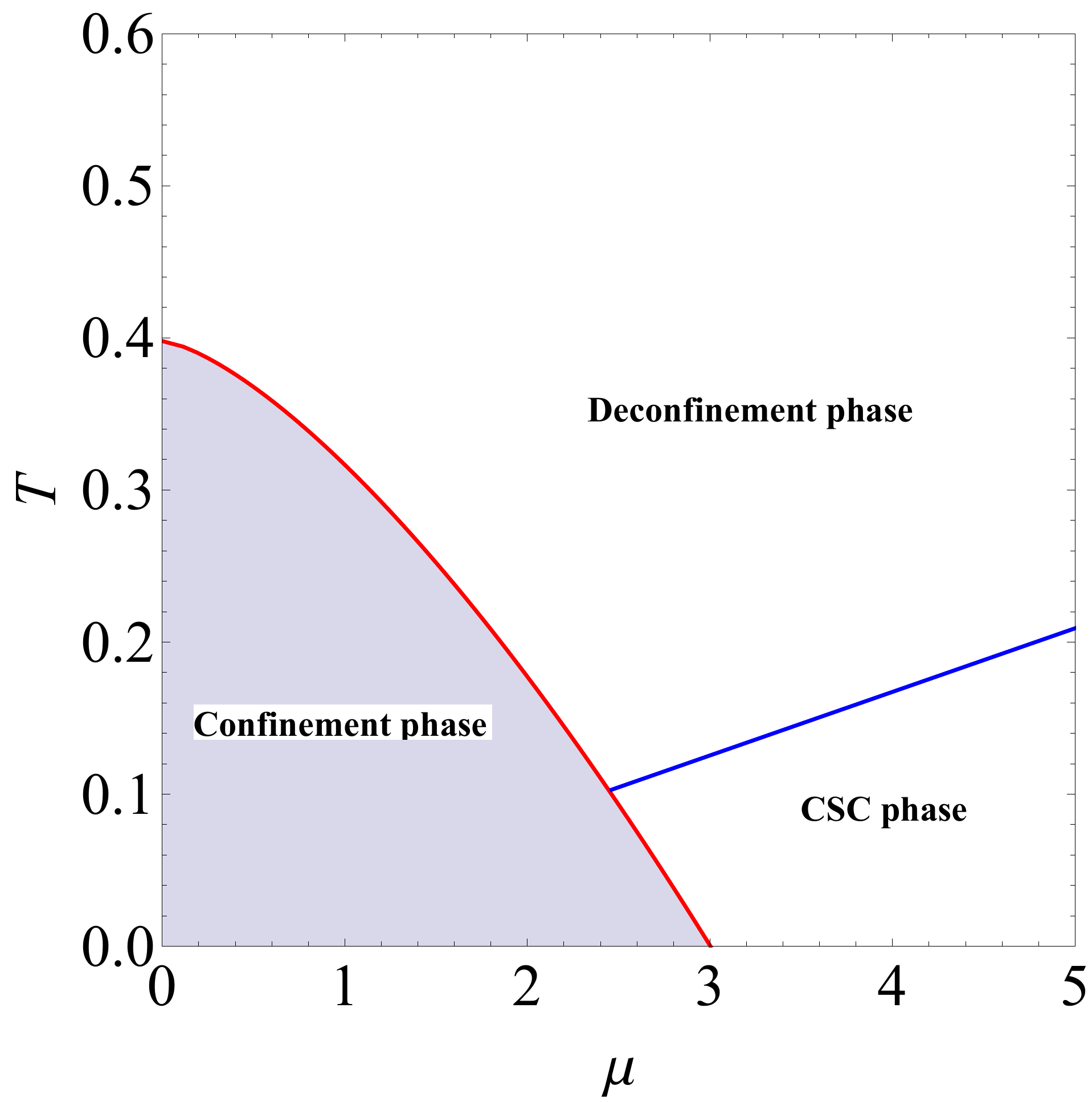}\\
\includegraphics[width=0.45 \textwidth]{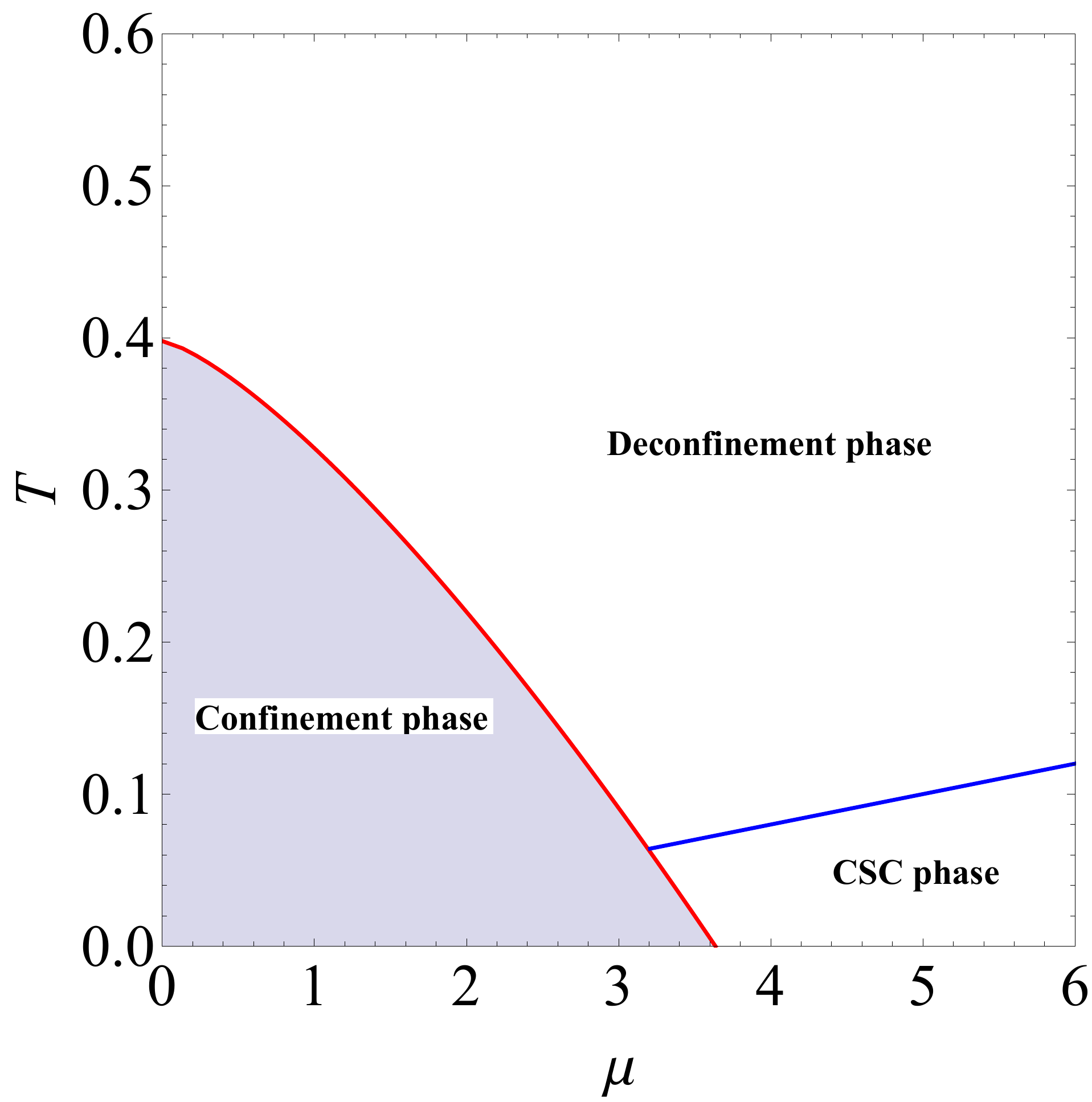}
\hspace*{0.05\textwidth}
\includegraphics[width=0.45 \textwidth]{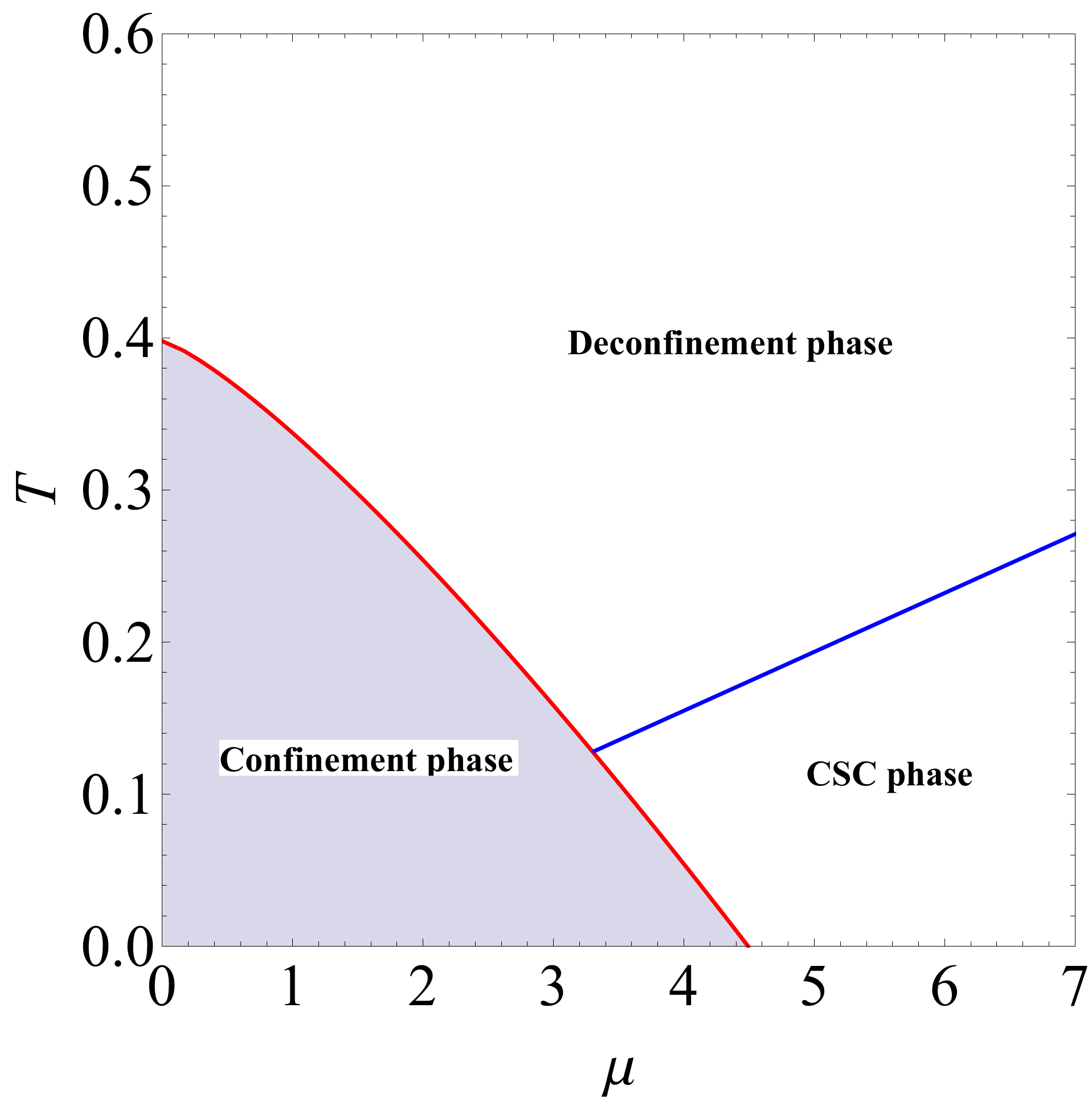}
\end{tabular}
  \caption{The phase diagram is shown for various values of the power parameter $s$ and the color number $N_c$. Top-left panel: $s=4/5$ and $N_c=2$. Top-right panel: $s=5/7$ and $N_c=2$. Bottom-left panel: $s=2/3$ and $N_c=3$. Bottom-right panel: $s=5/8$ and $N_c=3$.}\label{PDYMT-Nc23}
\end{figure}

From the results derived above, we can show the phase diagram of the present holographic model for the realistic YM theory in the $\mu-T$ plane for various values of the power parameter $s$ and the color number $N_c$ in Fig. \ref{PDYMT-Nc23}.
The red curve refers to the critical curve corresponding to the deconfinement/confinement phase transition. The region below the red curve corresponds to the confinement phase due to its gravitational dual being thermodynamically preferred. The region above the red curve and the blue line represents the (normal) deconfinement phase. The CSC phase appears in the region of the high chemical potential and the low temperature. Hence, the region below the blue line which refers to the critical line associated with the CSC phase transition represents the CSC phase. It is found that the region of the CSC phase becomes larger when the power parameter $s$ decreases, which corresponds to the growth of the slope of the  blue line as seen in Tables \ref{Nc2} and \ref{Nc3}. This suggests that the CSC phase is more stable in the regime of the small power parameter such that it is larger than $1/2$ and it should not lead to the appearance of the CSC state in the confinement phase. In particular, this phase diagram shows that the present holographic model with the power-law Maxwell field for the reasonable values of the power parameter $s$ can provide a gravitational dual description for the CSC phase transition of the realistic YM theory which appears in the deconfinement phase but not in the confinement phase.

\section{\label{sum} Summary}

We have studied the color superconductivity (CSC) phase transition in the realistic Yang-Mills theory (i.e. $N_c\geq2$ where $N_c$ is the color number of quarks) using a simple holographic model. This holographic model is constructed by Einstein gravity coupled minimally to the matter part which consists of a power-law Maxwell field and a complex scalar field which are dual to the baryon symmetry and the diquark operator, respectively. By analyzing the breaking of the Breitenlohner-Freedman bound and solving numerically the equations of motion in the configuration of the planar AdS black hole dual to the deconfinement phase, we have pointed out that the scalar field of the small charge ($q\leq1$) can condense around the event horizon of the planar AdS black hole if the power parameter characterizing for the power-law Maxwell field is sufficiently lower than one but above $1/2$. The condensation of the scalar field with the small charge in the gravitational dual theory means that the corresponding diquark operator develops a nonzero vacuum expectation value (VEV) and hence the CSC phase transition with the large color number ($N_c\geq2$) appears in the deconfinement phase, which is not found in the gravitational dual description with the usual Maxwell field. Furthermore, we have shown that the power parameter $s$ is not arbitrary below one (but above $1/2$) because when $s$ is sufficiently far away from one it leads to the condensation of the scalar field in the configuration of the AdS soliton or the nonzero VEV of the diquark operator in the confinement phase which would not be compatible with the nonzero VEV of the color nonsinglet operator. Finally, we have obtained the phase diagram in the plane of the temperature and the chemical potential.

The current work also claims that the condensate of the scalar field is actually easier to form for the smaller power parameter, which has been previously indicated in the investigation of the holographic superconductors with the power-law Maxwell field in the probe limit \cite{Pan2011,Salahi2016} and with the backreaction \cite{Jiang2016}. The critical values of the temperature and chemical potential grow with the decrease of the power parameter, as seen in Fig. \ref{PDYMT-Nc23}. However, there are main differences between these works and the current work. First, the gravitational dual model of the 4D usual superconductors is studied in five dimensions. Whereas, the gravitational dual description of the 4D color superconductors is considered in six dimensions with one dimension compactified on a circle $S^1$. This is because the boundary gauge field theory possesses a confinement scale which is identified with the inverse radius of the $S^1$. Second, besides studying the effect of the power-law Maxwell field on the critical temperature like in holographic superconductors, the current world studies the relevant effects on the color number of quarks (or the $U(1)_B$ charge of the condensate operator) and the critical chemical potential which are the important parameters to trigger the CSC phase transition. Third, unlike the holographic model of the usual superconductors which describes the low- and high-temperature phases corresponding to the superconducting and normal states, the holographic model of the color superconductor represents three different phases: they are the confinement, normal deconfinement, and CSC phases corresponding to the low $T$ and $\mu$, the high $T$ and high $\mu$, and the low $T$ and high $\mu$, respectively.

\end{document}